\pdfoutput=1
\documentclass{aa}  
\usepackage{graphicx}
\usepackage{siunitx}
\usepackage[breaklinks]{hyperref}
\usepackage{array}
\usepackage{upgreek}

\DeclareSIUnit\gauss{G}

\usepackage{url}
\usepackage{txfonts}
\graphicspath{{./figures/}}

\begin{document} 

   \title{Protostellar accretion traced with chemistry}
   \subtitle{Comparing synthetic C$^{18}$O maps of embedded protostars to real observations}

   \author{S. Frimann\inst{\ref{inst1}}
          \and
          J. K. J\o rgensen\inst{\ref{inst1}}
          \and
          P. Padoan\inst{\ref{inst2}}
          \and
          T. Haugb\o lle\inst{\ref{inst1}}
          }

   \institute{Centre for Star and Planet Formation, Niels Bohr Institute and Natural History Museum of Denmark, University of Copenhagen, \O ster Voldgade 5-7, DK-1350 Copenhagen K, Denmark \\ \email{sfrimann@nbi.ku.dk} \label{inst1}
   \and
   ICREA and Institut de Ci\`{e}ncies del Cosmos, Universitat de Barcelona, IEEC-UB, Mart\'{i} Franqu\`{e}s 1, E-08028 Barcelona, Spain \label{inst2}}

   \date{Received; accepted}
 
  \abstract
   {Understanding how protostars accrete their mass is a central question of star formation. One aspect of this is trying to understand whether the time evolution of accretion rates in deeply embedded objects is best characterised by a smooth decline from early to late stages or by intermittent bursts of high accretion.}
   {We create synthetic observations of deeply embedded protostars in a large numerical simulation of a molecular cloud, which are compared directly to real observations. The goal is to compare episodic accretion events in the simulation to observations and to test the methodology used for analysing the observations.}
   {Simple freeze-out and sublimation chemistry is added to the simulation, and synthetic C$^{18}$O line cubes are created for a large number of simulated protostars. The spatial extent of C$^{18}$O is measured for the simulated protostars and compared directly to a sample of 16 deeply embedded protostars observed with the Submillimeter Array. If CO is distributed over a larger area than predicted based on the protostellar luminosity, it may indicate that the luminosity has been higher in the past and that CO is still in the process of refreezing.}
   {Approximately \SI{1}{\percent} of the protostars in the simulation show extended C$^{18}$O emission, as opposed to approximately \SI{50}{\percent} in the observations, indicating that the magnitude and frequency of episodic accretion events in the simulation is too low relative to observations. The protostellar accretion rates in the simulation are primarily modulated by infall from the larger scales of the molecular cloud, and do not include any disk physics. The discrepancy between simulation and observations is taken as support for the necessity of disks, even in deeply embedded objects, to produce episodic accretion events of sufficient frequency and amplitude.}
  {}


   \maketitle
%

\newcommand{\ramses}{\texttt{RAMSES}\xspace}
\newcommand{\radmc}{\texttt{RADMC-3D}\xspace}
\newcommand{\co}{C$^{18}$O\xspace}

\newcommand{\sfrimann}[1]{\textcolor{red}{\textbf{(#1)}}}
\newcommand{\troels}[1]{\textcolor{blue}{\textbf{(#1)}}}

\section{Introduction}
\label{sec:introduction}

A protostar obtains most of its mass during the early embedded stages, where it is surrounded by large amounts of dust and gas. At this stage the protostellar luminosity is dominated by the release of gravitational energy due to accretion onto the central object. Numerous large-scale surveys (e.g. \citealt{Evans:2009bk,Kryukova:2012ea,Dunham:2013do}) have shown that the distribution of protostellar luminosities (the protostellar luminosity function, PLF) is a wide distribution, spanning more than three orders of magnitude, with a median luminosity of 1\,$L_\sun$--2\,$L_\sun$. To form a solar mass star within the duration of the embedded stage of \SI{\sim 0.5}{Myr} \citep{Evans:2009bk}, an average accretion rate of \num{2e-6}\,$M_\sun \, \mathrm{yr}^{-1}$ is needed. Assuming a protostellar mass and radius of 0.5\,$M_\sun$ and 2.5\,$R_\sun$, this implies an accretion luminosity of $\sim$10\,$L_\sun$, which is an order of magnitude above the observed median. The situation is further aggravated by the fact that other sources of luminosity, such as deuterium burning and Kelvin-Helmholtz contraction, also have to be taken into consideration.

The inability of simple physical arguments to accurately predict the observed PLF is known as the luminosity problem and was first noticed by \citet{Kenyon:1990ki}. Since its discovery more than two decades ago much work has gone into reconciling theory and observations; the problem has been approached through both analytical modelling (e.g. \citealt{Dunham:2010bx,Offner:2011ex,Myers:2012ea}) and numerical simulations (e.g. \citealt{Dunham:2012ic,Padoan:2014ho}). A variety of different accretion scenarios, which can reproduce both spread and median of the observed PLF, have been presented, indicating that reproduction of the PLF alone is not sufficient to discriminate between different accretion scenarios.

The preferred scenario for resolving the luminosity problem requires a gradual decrease in the mean accretion rate together with episodic accretion bursts. Such accretion bursts are thought to be modulated by variations in the infall of material from large scales \citep{Padoan:2014ho} and by instabilities in circumstellar disks \citep{Bell:1994cp,Armitage:2001jl,Vorobyov:2005kv,Zhu:2009fv,Martin:2011jz}. Additionally, evidence from simulations indicates that episodic accretion may be needed to reproduce the peak of the initial mass function (IMF; \citealt{Lomax:2014bw}). Bursts are well-established in pre-main-sequence stars where they appear as FU Orionis type objects, which are young stellar objects showing strong, long-lived optical luminosity bursts (\citealt{Herbig:1977gf}; see \citealt{Audard:2014gb} for a recent review). It is not known whether deeply embedded protostars are subject to episodic outbursts to the same degree as their older, more evolved counterparts, since searches for luminosity bursts are most easily performed at optical and near-infrared wavelengths where deeply embedded objects cannot be detected. However, recently \citet{Safron:2015ew} reported on the detection of the outbursting Class~0 protostar, HOPS~383, detected through brightening in the mid-infrared.

\begin{figure*}
\sidecaption
\includegraphics[width=12cm]{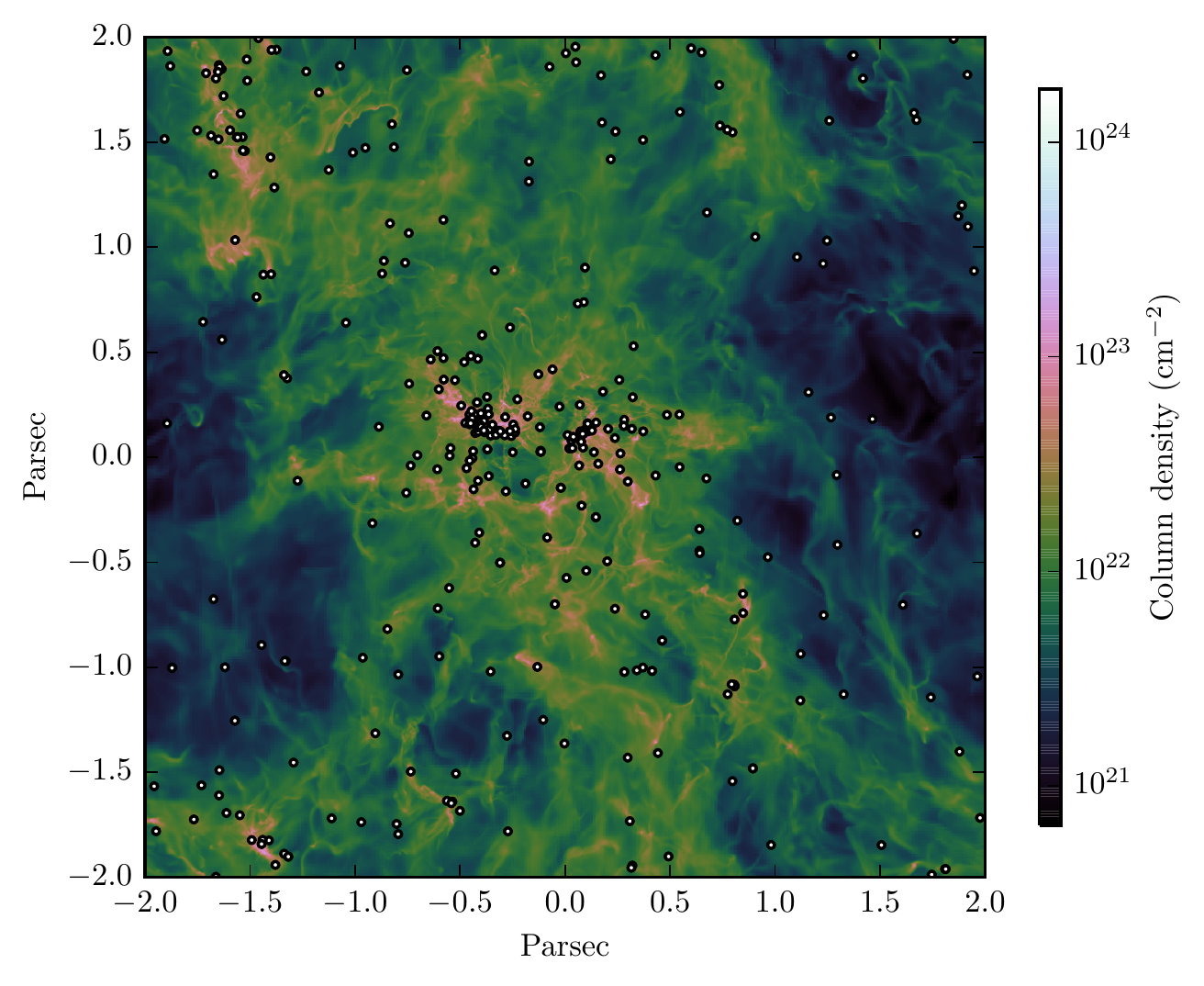}
   \caption{Projected gas number density and sink particle positions of a snapshot in the simulation \SI{2.2}{Myr} after the formation of the first sink. The total number of sink particles at this point is 454.}
   \label{fig:densitymap}
\end{figure*}

An accretion burst heats up the dust in the surrounding envelope on timescales that are negligible compared to the typical duration of an accretion burst \citep{Johnstone:2013je}. This has profound effects on the chemistry in the envelope since, following the increase in temperature, molecular ices will begin to sublimate from the dust grains and move to the gas-phase. The envelope cools rapidly once the accretion burst ends, and the molecules begin to refreeze back onto the dust grains. The process of refreezing proceeds slowly, meaning that molecules remain in the gas-phase for several thousand years following the end of the accretion burst and can therefore be used to determine whether the protostar has had a larger luminosity in the past (e.g. \citealt{Visser:2012dp,Vorobyov:2013kq,Visser:2015ew}).

One of the best examples to date of using chemistry as evidence for a past luminosity burst is the case of IRAS~15398--3359, which was found by \citet{Jorgensen:2013ej} to have H$^{13}$CO$^+$ distributed in a ring-like structure surrounding the protostar. The best explanation for the lack of H$^{13}$CO$^+$ in the innermost regions is for it to have been destroyed by water, except that the temperatures in the central regions of the envelope are too low for water ice to sublimate. The authors conclude that for water to be present in the gas-phase at these temperatures, the protostar must have undergone an accretion burst within the past \SIrange{100}{1000}{yr}.

An alternative approach was presented by \citet{Jorgensen:2015kz}, who measured the spatial extent of \co around 16 embedded protostars, using observations from the Submillimeter Array (SMA). By comparing the measured extents with the protostellar luminosities they found that approximately half the protostars in the sample have \co distributed over a larger area than can be explained by their current luminosity, which is attributed to the protostars having had a significantly higher luminosity sometime during the past few thousand years.

This paper presents synthetic observations of \co for a large sample of deeply embedded protostars, which are subjected to the same kind of analysis as in \citet{Jorgensen:2015kz} and compared directly to their results. The synthetic observations are created from a large numerical simulation of a molecular cloud, which encompasses simultaneously protostellar system and molecular cloud scales. Simple chemistry is added to the model in the post-processing stage to be able to follow the sublimation and freeze-out of CO.

There are two goals of comparing synthetic and real observations: (1) to test the methodology employed for measuring the spatial extent of \co in the observations and see how well it performs at capturing past bursts, and (2) to study how the spatial extent of \co, measured from the simulation, compares to the extents measured in the observations. The second point is of particular interest since the simulation is a rerun of the simulation presented in \citet{Padoan:2014ho}, where it was demonstrated that a gradual decrease in the accretion rate, together with accretion variability due to infall of material from large scales, can reproduce the observed PLF. The question at hand is whether this is enough to also explain other observables that are influenced by the accretion history, such as the measured \co extents. The outline of the paper is as follows: Section~\ref{sec:methods} introduces the numerical simulation as well as the radiative transfer post-processing used to create synthetic observables. Section~\ref{sec:synthobs} presents the analysis applied to the synthetic observations and compare them to the observations. The results show that, while the methodology is capable of at measuring the extent of the gas-phase distribution of CO, the extents measured in the simulation are not as broadly distributed as in the observations. Section~\ref{sec:discussion} discusses the reasons and implications of this discrepancy, while Sect.~\ref{sec:Summary} summarises the findings of the paper.

\section{Methods}
\label{sec:methods}

\subsection{The numerical simulation}
\label{sec:simulation}

The numerical simulation has been run using the adaptive mesh refinement (AMR) code, \ramses \citep{Teyssier:2002fj}, modified to include random turbulence driving, a novel algorithm for sink particles, and technical improvements allowing for efficient scaling to several thousand cores. The simulation is part of a suite of simulations run by Haugb\o lle et al.~in preparation, investigating the IMF. It is simultaneously a rerun of the simulation presented by \citet{Padoan:2014ho}, except for a few aspects described below, and we refer to that paper for details on the numerical setup. The simulation encompasses simultaneously molecular cloud and protostellar system scales with a box-size of \SI{4}{pc} and a minimum cell size of \SI{50}{AU}. The root grid of the simulation is $256^3$, with an additional 6 levels of AMR on top. The refinement is based on density and is such that the Jeans length is resolved with at least 14 cells everywhere. The total mass in the box is 2998\,$M_\sun$ corresponding to an average number density of \SI{795}{cm^{-3}} assuming a molecular weight of 2.4. The initial magnetic field has a strength of \SI{7.2}{\micro\gauss}. The simulation is run for a total time of \SI{2.6}{Myr} and forms a total of \num{635} sink particles. Figure~\ref{fig:densitymap} shows the projected density, along with the sink particles, of one of the snapshots in the simulation.

There are two aspects where the present run differs from that of \citet{Padoan:2014ho}: (1) a slightly improved MHD solver and (2) changes in the numerical sink particle parameters. In the current run the slope reconstruction and limiting procedure interpolating from cell centres to cell interfaces is more isotropic and less diffusive, while still being stable in supersonic flows. Both the original and the present run use the HLLD solver, but switch to a diffusive local Lax Friedrich (LLF) solver at trouble cells. In the original run we changed to the LLF solver at cells where the local fast mode speed exceeded 100 times the local sound speed, while for the present run we were able to switch at 300, reducing the typical number of cells where LLF is employed to \SI{\sim 2}{\percent}. The sink particle parameters were changed slightly. The density threshold for sink particle creation is increased with a factor of 13 to \num{2.1e6} times the average density or \SI{1.7e9}{cm^{-3}}. This makes the algorithm more robust and we can reduce the minimum distance between new and existing sink particles from 16 to 8 cells, corresponding to \SI{400}{AU}. We also reduced the threshold density for considering gas accretion with a factor of 30 to an overdensity of 5.3 or \SI{4226}{cm^{-3}}, to make it possible to follow accretion in evolved systems down to \num{e-10}\,$M_\sun$\,yr$^{-1}$. The combined effect of the changes is less diffusive gas dynamics and an almost perfect suppression of spurious sink particles. We denote sink particles spurious if they are created by chance close to other sink particles instead of from bona fide gravitationally collapsing regions of the gas. This can happen, for example, in the accretion flow close to an existing sink particle, where the gas density can get very high. While it does not affect the conclusions of the \citeauthor{Padoan:2014ho} paper, we later found a method to estimate the fraction of spurious sinks and found it to be \SI{\sim 10}{\percent} of all sinks in the original simulation. In the rest of the paper we refer to the sink particles as ``protostars''.

\subsection{Post-processing}
\label{sec:postprocess}

To produce observables the simulation is post-processed using the radiative transfer code \radmc\footnote{\url{http://www.ita.uni-heidelberg.de/~dullemond/software/radmc-3d/}} (see \citealt{Dullemond:2004iy} for a description of the two-dimensional version of this code). Much of the post-processing is equivalent to what was done in \citet{Frimann:2015}, so the reader is referred to that paper for details. The most important aspects of the post-processing are repeated here, along with a full description of the synthetic line observations.

As in \citet{Frimann:2015}, \radmc is run on cubical cut-outs with side lengths of \SI{\approx 30000}{AU}, centred around individual protostars. \radmc can handle AMR grids natively and it is therefore not necessary to resample the simulation output onto a regular grid. The cut-outs are made by cycling through all protostars in all snapshots of the simulation, with each cut-out corresponding to one \radmc model. Each individual cut-out may contain several protostars, but for each \radmc model the central protostar is the only source of luminosity.

\subsubsection{Radiative transfer}

The thermal Monte Carlo module  of \radmc uses the method of \citet{Bjorkman:2001du} to calculate the dust temperatures in the models. The method relies on the propagation of a number of ``photon packets'' through the model, which, in our case, has been set to ten million. We use the OH5 dust opacities of \citet{Ossenkopf:1994tq}, corresponding to coagulated dust grains with thin ice mantles at a density of $n_\mathrm{H_2}$\,$\sim$\,\SI{e6}{cm^{-3}}, which have been found to be appropriate for dense cores by several studies (e.g. \citealt{Shirley:2002co,Shirley:2011jz}). A dust-to-gas mass ratio of 1:100 is assumed everywhere. The luminosity of the central protostar is calculated as the sum between the accretion luminosity, arising from accretion onto the protostar, and the photospheric luminosity, arising from deuterium burning and Kelvin-Helmholtz contraction
\begin{equation}
L_\star = L_\mathrm{phot} + L_\mathrm{acc} = L_\mathrm{phot} + f_\mathrm{acc} \frac{G\dot{m}m_\star}{r_\star},
\label{eq:Lstar}
\end{equation}
where $m_\star$ is the mass of the protostar; $\dot{m}$ the instantaneous accretion rate onto the protostar; $r_\star$ the protostellar radius, taken to be \num{2.5}\,$R_\sun$; and $f_\mathrm{acc}$ is the fraction of accretion energy radiated away, taken to be 1. The accretion rate, $\dot{m}$, is calculated from an auxiliary output file of the simulation, recording the time evolution of the protostellar masses. The sampling rate of this time series has a median value of \SI{\approx 30}{yr} meaning that accretion rates are averaged over this time interval. $L_\mathrm{phot}$ is calculated using the pre-main-sequence tracks of \citet{DAntona:1997vs}, where we follow \citet{Young:2005ic} and add \SI{100}{kyr} to the tabulated ages to account for the time difference between the beginning of core-collapse and the onset of deuterium burning. The protostars are assumed to emit as perfect black bodies with an effective temperature of \SI{1000}{K}.

\subsubsection{Freeze-out and sublimation chemistry}
\label{sec:freezeoutsub}

Simple freeze-out and sublimation chemistry is added to the code to simulate the situation following an accretion burst where molecules are still in the process of refreezing back onto the dust grains. Following \citet{Rodgers:2003dg} the thermal sublimation rate of ices sitting on top of dust grains can be written as
\begin{equation}
  k_\mathrm{sub} = \nu \, \exp \left( -\frac{E_b}{k_BT_\mathrm{dust}} \right). \label{eq:ksub}
\end{equation}
The prefactor, $\nu$, is set to \SI{2e12}{s^{-1}}, appropriate for a first order reaction \citep{Sandford:1993da}. $E_b$ is the binding energy of the molecule on the dust grains. We adopt a value for the binding energy, $E_b/k_B$, of \SI{1307}{K} appropriate for a mixture between water and CO ice \citep{Noble:2012hq}. This corresponds to a sublimation temperature, $T_\mathrm{sub}$, of approximately \SI{28}{K}, where $T_\mathrm{sub}$ is defined as the temperature where the sublimation timescale is one year.

Following again \citet{Rodgers:2003dg}, the freeze-out rate can be written as
\begin{equation}
  k_\mathrm{dep} = \SI{2.88e-11}{s^{-1}} \sqrt{\frac{T_\mathrm{gas}}{\mu \, \SI{10}{K}}} \, \frac{n_{\mathrm{H}_2}}{\SI{e6}{cm^{-3}}}, \label{eq:kdep}
\end{equation}
where $\mu$ is the molecular weight of the considered species and the prefactor has been calculated assuming $n_\mathrm{grain}/n_{\mathrm{H}_2}$\,=\,\num{2e-12}, an average dust grain radius of \SI{0.1}{\micro\metre}, and unit sticking coefficient. Gas and dust temperatures are assumed to be equal.

The goal is to be able to calculate the gas-phase abundance of CO, $X_\mathrm{co,gas} \equiv n_\mathrm{co,gas}/n_{\mathrm{H}_2}$, in each cell of the \radmc model, while taking into account the luminosity history of the central protostar. The total CO abundance, $X_\mathrm{co,tot} = X_\mathrm{co,gas} + X_\mathrm{co,grain}$, is set to the canonical value of \num{e-4}. \ramses is a grid-based model, which means that it is not possible to follow the dynamical evolution of individual particles -- a follow-up study with tracer particles included in the \ramses simulation will be able to do this, but probably not for a large sample of objects. Instead, the gas-phase abundance of CO is calculated using the procedure sketched out in the following paragraph.

The density distribution in the \radmc model is held fixed, while the temperatures in the model are recalculated for a range of different luminosities, based on the luminosity history of the central protostar. The temperatures are not recalculated for each unique luminosity the protostar has had in the past, but are sampled on a logarithmic luminosity grid starting from \num{e-5}\,$L_\sun$ and increasing in steps of \SI{3}{\percent} from there. Naturally, the temperatures are only calculated for grid points within the luminosity range covered by the protostar. The grid is interpolated using a power law from the nearest grid point, $T_\mathrm{interpolated}/T_\mathrm{grid} = (L_\mathrm{actual}/L_\mathrm{grid})^{0.2}$, where the power law exponent has been measured empirically. $X_\mathrm{co,gas}$ is calculated individually in each cell of the \radmc model, effectively making each cell its own one-zone model. If the dust temperature in a cell is higher than the sublimation temperature, $T_\mathrm{dust} > T_\mathrm{sub}$, then $X_\mathrm{co,gas} = X_\mathrm{co,tot}$; on the other hand if $T_\mathrm{dust} < T_\mathrm{sub}$ then $X_\mathrm{co,gas}$ is calculated using Eq.~\eqref{eq:kdep}, with $\min (X_\mathrm{co,gas}) = 0.01 \times X_\mathrm{co,tot}$ to reflect the fact that non-thermal desorption processes (e.g. cosmic rays) ensure that CO is never completely frozen-out. The system is evolved starting from the maximum luminosity it has had in the past, where $X_\mathrm{co,gas}$ is initialised so that $X_\mathrm{co,gas} = X_\mathrm{co,tot}$ for $T_\mathrm{dust} > T_\mathrm{sub}$, and $X_\mathrm{co,gas} = 0.01 \times X_\mathrm{co,tot}$ for $T_\mathrm{dust} < T_\mathrm{sub}$.

\subsubsection{Synthetic line cubes}

\begin{figure*}
\includegraphics[width=18cm]{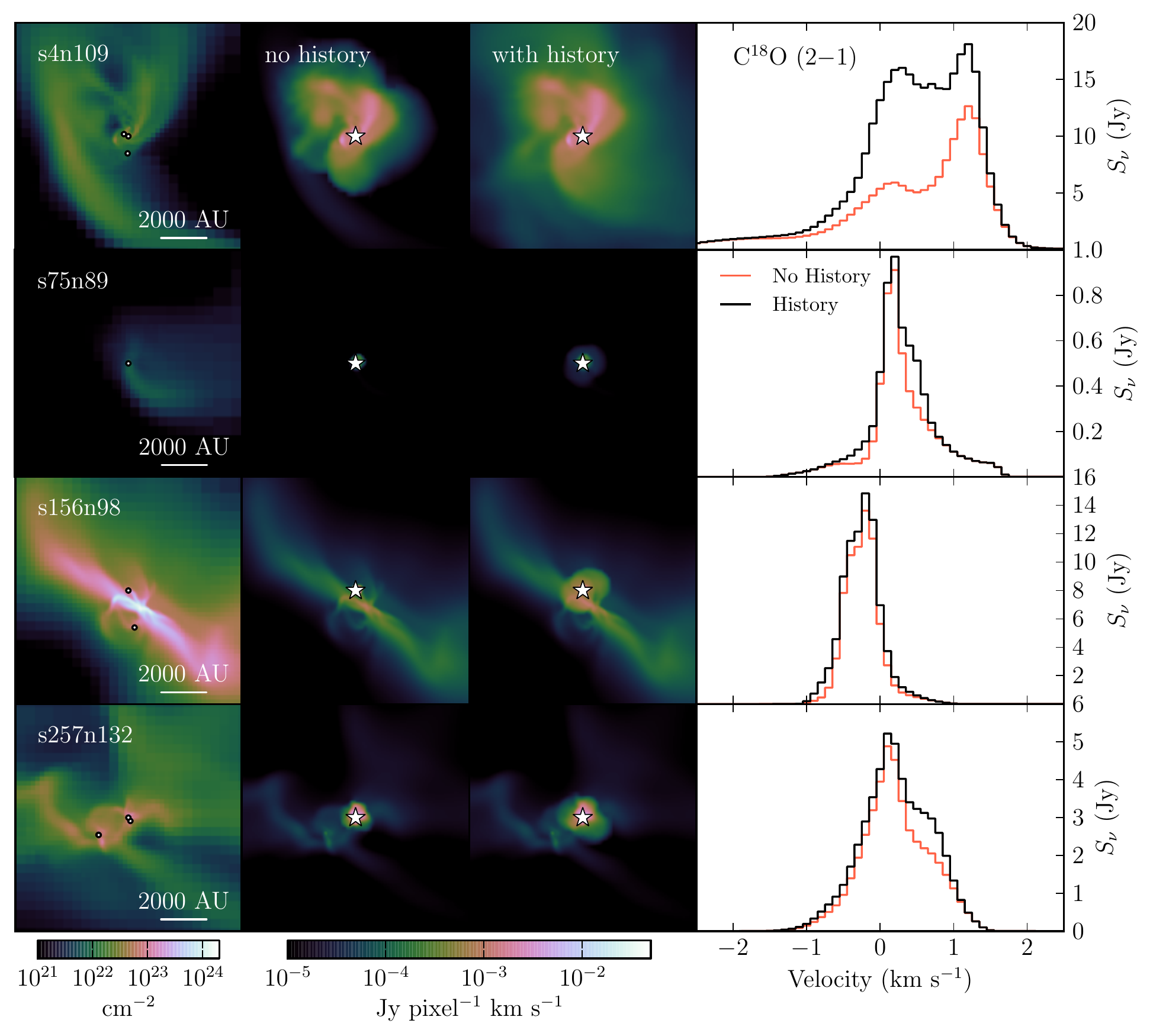}
\caption{From raw simulation to synthetic observables for four different systems. From left to right: projected gas column density of the raw simulation, with dots indicating protostars; zero moment maps of the \co 2--1 line; line spectra of the two zero moment maps. In the column labelled ``no history'' the \co number densities have been calculated without taking the luminosity history of the central protostar into consideration, while, in the column labelled ``history'', the luminosity history has been included. The line spectra have been calculated by integrating over the entire field of view as seen in the moment maps.}
\label{fig:synthobs}
\end{figure*}

The end product of the post-processing are synthetic observables, in the form of line cubes. The main isotopolouge of CO is optically thick at the densities found in protostellar cores so an optically thin isotopolouge is typically used. In this case, the synthetic line cubes are of the \co~$J$\,=\,2--1 line and are produced using the line radiative transfer module in \radmc assuming Local Thermodynamic Equilibrium (LTE). \citet{Jorgensen:2002ci} showed that, for low $J$ rotational transitions and the high densities characteristic of deeply embedded objects, LTE is a good approximation. The gas-phase abundances of CO are calculated as described in Sect.~\ref{sec:freezeoutsub} above and we calculate the number densities of \co assuming a $^{16}$O/$^{18}$O ratio of \num{500}. The spatial extent covered by the cubes is \SI{10000x10000}{AU} and they are given a spatial resolution of \SI{25}{AU}. This is a factor of two higher than the maximum resolution of the simulation, as well as in the \radmc model, and we choose this slight over-sampling to ensure that the resolution of the model is not accidentally degraded by ray tracing along cell borders. The width of the spectral window is \SI{10}{km.s^{-1}}, with a resolution of \SI{0.1}{km.s^{-1}} centred around the \co 2--1 line. Figure~\ref{fig:synthobs} shows examples of zero moment maps of some of the synthetic line cubes along with spectra of the \co 2--1 line and column density maps of the raw simulation. The column labelled ``no history'' shows zero moment maps calculated from \radmc models, where the gas-phase abundance of CO only depends on the current luminosity; in the column labelled ``history'', the luminosity history of the protostar has been taken into account. All results and figures presented in this paper are derived assuming a source distance of \SI{235}{pc}.

\begin{table*}
\caption{Example object parameters. The first column is the protostellar mass, followed by its age and luminosity. $T_\mathrm{bol}$ is the bolometric temperature as defined by \citet{Myers:1993en}. $M_{10\,000\,\mathrm{AU}}$ is the gas mass within a radius of \SI{10000}{AU} from the protostar. Finally, $n_0$ and $p$ are the fitted parameters for the power law, $\rho(r) = n_0 (r/\SI{100}{AU})^{-p}$.}
\label{tbl:object}
\centering
\begin{tabular}{l c c c c c c c}     
\hline\hline
 & $M_\star$ ($M_\sun$) & Age (kyr) & Luminosity ($L_\sun$) & $T_\mathrm{bol}$ (K) & $M_{10\,000\,\mathrm{AU}}$ ($M_\sun$)\tablefootmark{a} & $n_0$ (\si{cm^{-3}}) & $p$ \\
\hline
s4n109   & 5.1 & 1037 & 70.4 & 259 & 0.5 & \num{1.6e6} & 1.5 \\
s75n89   & 0.2 &  248 & 0.9  & 330 & 0.1 & \num{5.5e5} & 1.3 \\
s156n98  & 0.1 &   16 & 1.4  &  54 & 2.4 & \num{8.4e6} & 1.3 \\
s257n132 & 0.3 &  175 & 5.6  &  84 & 1.0 & \num{5.2e6} & 1.4 \\
\hline
\hline                  
\end{tabular}
\tablefoot{
\tablefoottext{a}{Gas mass inside a radius of \SI{10000}{AU}.}
}
\end{table*}

\section{Synthetic observations}
\label{sec:synthobs}

The synthetic observations analysed in this paper are compared to the observations presented by \citet{Jorgensen:2015kz}. To create a context for the later comparison, these observations are described briefly here. The authors use the SMA to observe the \co~$J$\,=\,2--1 line towards 16 deeply embedded Class~0 and~I sources in a number of the nearby star-forming regions. The observations utilise SMA's compact configuration, which gives a spatial resolution of roughly \ang{;;2}, corresponding to a baseline radius of approximately 50\,k$\uplambda$. Resolved zero moment maps, of the \co emission, show centrally concentrated emission with evidence of some low surface brightness extended emission for a few objects. The authors measure the deconvolved extent of CO emission by fitting one-dimensional Gaussians to the $(u,v)$-data and find, after allowing for various observational uncertainties, that approximately half of the protostars in the sample have CO distributed over a larger area, than can be explained by the current protostellar luminosity, indicating that the luminosity has been larger sometime in the past.

\subsection{Target selection}
\label{sec:targetselection}

Over the course of its running time, the simulation forms a total of \num{635} protostars and stores \num{119} snapshots, yielding a total of \num{30340} \radmc models (this number is not equal to the total number of protostars times the number of snapshots because not all protostars are present in all snapshots). It is impractical to calculate molecular line cubes for such a large number of models, both in terms of computing time and space needed to store the cubes. Additionally, the purpose of this study is to compare the synthetic cubes with the 16 deeply embedded protostars observed by \citet{Jorgensen:2015kz}, hence only objects that match the observed sample should be included. For these reasons a number of selection criteria are applied to the full sample of \radmc models, with the goal of reducing the sample to only include the objects of interest. These selection criteria are discussed below, but the end result is that the sample of \radmc models is reduced from \num{30340} to \num{1952}. In addition to this, for each \radmc model in the sample, three synthetic line cubes are calculated for three orthogonal viewing angles, effectively increasing the sample size with a factor of three.

The first selection criterion removes all sources with a luminosity outside the range 0.5\,$L_\sun$--200\,$L_\sun$. The motivation for this criterion is to only include sources where CO is expected to be in the gas-phase on the spatial scales which can be probed by the interferometer. The second selection criterion seek to avoid binary stars by requiring that the distance to the nearest protostar is >\,\SI{400}{AU}. The remaining selection criteria all seek to pick out the deeply embedded objects. The first of these is based on the SEDs of the protostars. Using the same method as in \citet{Frimann:2015} we calculate the bolometric temperature \citep{Myers:1993en} for all objects in the simulation and choose only objects with $T_\mathrm{bol}$\,<\,\SI{650}{K}, corresponding to Class~0 and~I protostars \citep{Chen:1995eo}. To select only objects with a strong central concentration of density, power law functions, of the form $\rho(r) = n_0 (r/\SI{100}{AU})^{-p}$, are fitted to the protostellar envelopes. To include the protostar in the sample, we require the fitted values, $n_0$ and $p$ to be >\,\SI{e5}{cm^{-3}} and~>\,\num{1.3} respectively. Finally, it is required that the total gas mass within a radius of \SI{10000}{AU} from the protostar is larger than $0.1\,M_\sun$. In this way we seek to match the sample of deeply embedded protostars observed by \citet{Jorgensen:2015kz}.

Figure~\ref{fig:synthobs} shows the projected gas density, zero moment maps and \co line spectra of four of the objects in the reduced sample. These objects are chosen so that they span the full range of luminosities in the sample and so that each object has had a larger luminosity in the past, making a visible difference on the moment maps whether or not the luminosity history is taken into consideration. Additional information about the four objects can be found in Table~\ref{tbl:object}.

\subsection{Analysis}
\label{sec:analysis}

\begin{figure}
\includegraphics[width=\hsize]{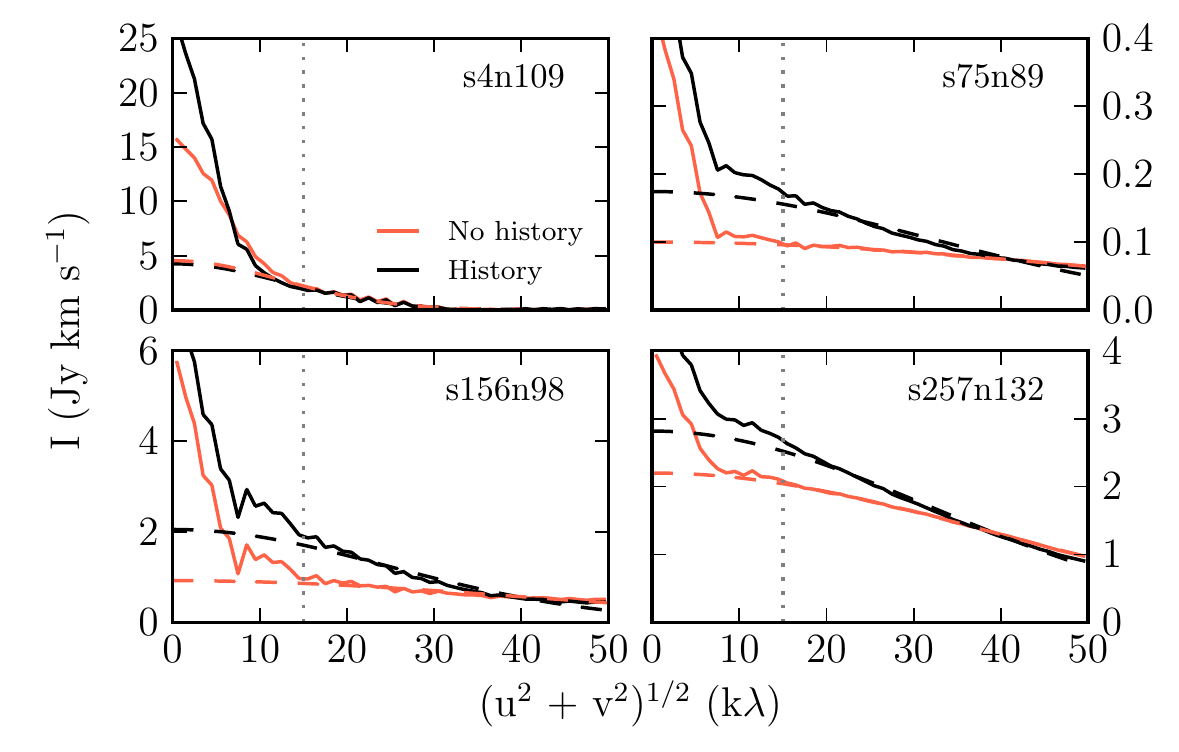}
   \caption{$(u,v)$-amplitude plots of the four objects shown in Fig.~\ref{fig:synthobs}. As before, the black line shows the situation including luminosity history and the red line the situation without. The dashed lines show the fitted Gaussians for each situation. The dotted line show the 15\,$\mathrm{k}\uplambda$ lower limit imposed on the fit. The upper limit on the fit is 50\,$\mathrm{k}\uplambda$.}
   \label{fig:uvamp}
\end{figure}

To measure the spatial extent of \co emission around the protostars in the sample, the synthetic line cubes are first multiplied with a primary beam and Fourier transformed so that the images are sampled in the $(u,v)$-plane for better comparison with interferometric observations. The $(u,v)$-data are subsequently fitted with a Gaussian to measure the deconvolved extent of \co emission (see Fig.~\ref{fig:uvamp} for some examples). \citet{Jorgensen:2015kz} did not use baselines $< 15\,\mathrm{k}\uplambda$ to avoid emission from the ambient medium, for example caused by external heating from the interstellar radiation field (ISRF). We do not include an interstellar radiation field, or similar, in our models, but retain the $15\,\mathrm{k}\uplambda$ limit for consistency with the observations. Likewise, baselines >50\,$\mathrm{k}\uplambda$ are disregarded to emulate the baseline coverage of the SMA.

\begin{figure}
\includegraphics[width=\hsize]{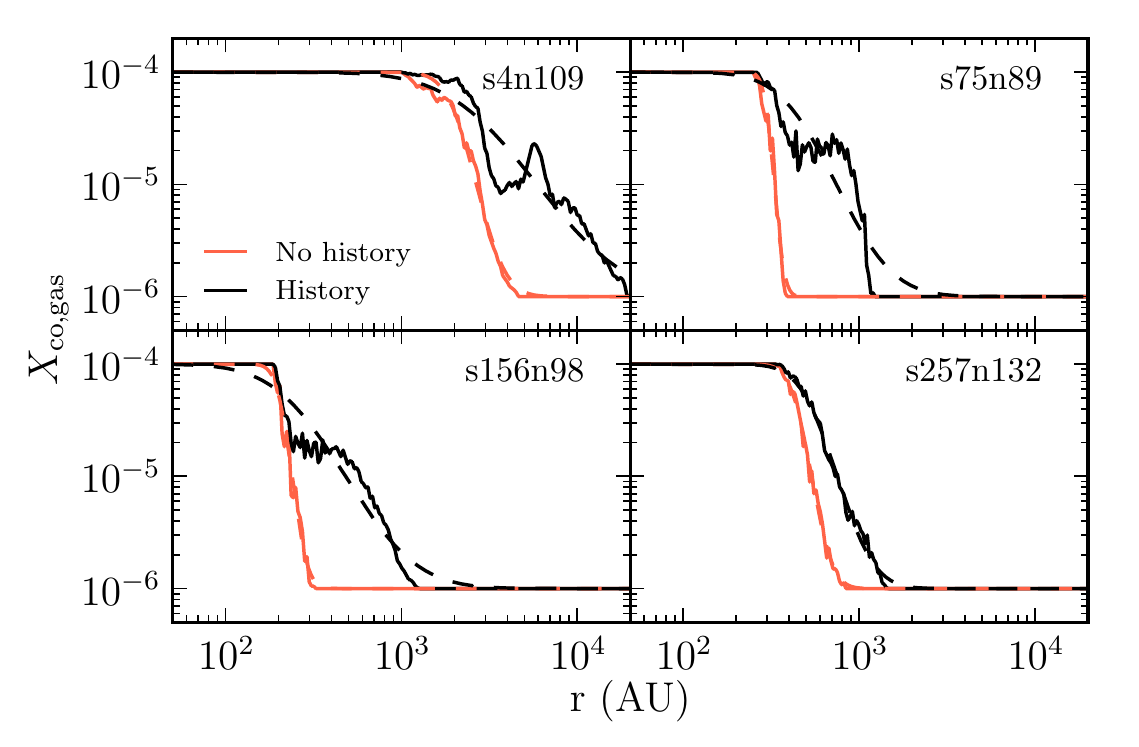}
   \caption{Abundance profiles of the four example objects. The profiles are fitted with cumulative Gaussian functions in loglog-space, shown as dashed lines.}
   \label{fig:abund}
\end{figure}

One of the features of Fourier transforms is that the Fourier transform of a Gaussian is itself a Gaussian. This means that, after fitting the Gaussian and finding the FWHM of the emission in the $(u,v)$-plane, the expected FWHM of the emission in the image plane can be found immediately by using the transformation
\begin{equation}
  \mathrm{FWHM}(\mathrm{arcsec}) = \frac{182}{\mathrm{FWHM}(\mathrm{k}\uplambda)}. \nonumber
\end{equation}
This expression can also be used to estimate the limits on the size-scales that can be probed with the interferometer, given the baseline limits of 50\,$\mathrm{k}\uplambda$ and 15\,$\mathrm{k}\uplambda$. A FWHM in the $(u,v)$-plane of 100\,$\mathrm{k}\uplambda$ ($2 \times 50$\,$\mathrm{k}\uplambda$) correspond to a FWHM in the image plane of \ang{;;1.8}, while a FWHM of 30\,$\mathrm{k}\uplambda$ ($2 \times 15$\,$\mathrm{k}\uplambda$) correspond to \ang{;;6.1}.

The fitted Gaussians are one-dimensional, which is expected to work well if the emission is roughly axisymmetric. To test if this is the case we also fitted two-dimensional Gaussians to the objects in the sample. On average, the minor and major axes of the two-dimensional Gaussians are within \SI{20}{\percent} of each other, with higher luminosities, where the emission is spread out over a larger area, having a higher probability of being asymmetric. Comparing the geometric mean of the minor and major axes of the two-dimensional Gaussians to the FWHM of the one-dimensional Gaussians reveals that, for \SI{90}{\percent} of the objects, they agree within \SI{1}{\percent}. This shows that, even in cases where the emission may be somewhat asymmetric, the one-dimensional fit will converge towards a reasonable value.

In addition to fitting Gaussians in the $(u,v)$-plane we also calculate abundance profiles of all objects by averaging $X_\mathrm{co,gas}$ over concentric shells around the protostar (see Fig.~\ref{fig:abund} for some examples). The abundance profiles make it possible to measure the spatial extent of the gas-phase CO independently of the fitted Gaussians. Specifically, the spatial scale of gas-phase CO is measured from the abundance profile by fitting a cumulative Gaussian function (an error function) in loglog-space to get the mean. The extents measured by the two methods differ by a roughly constant scale factor, which we measure to be \num[separate-uncertainty]{2.0\pm 0.2} (one-sigma uncertainty). The measurement has been limited to luminosities between 4\,$L_\sun$ and~10\,$L_\sun$ to avoid uncertainties due to the baseline limits. The results from the two methods can be compared directly, either by applying the scale factor or by comparing ratios between measurements.

\begin{table*}
\caption{Measured extents.}
\label{tbl:extents}
\centering
\begin{tabular}{l c c c c c c c c c c}     
\hline\hline
 & \multicolumn{3}{c}{$R_\mathrm{CO}$ (AU)} & Ratio\tablefootmark{a} & \multicolumn{2}{c}{Abund.\ radius (AU)} & Ratio\tablefootmark{a} & $\Delta t_\mathrm{peak}$ (kyr)\tablefootmark{b} & $L_\mathrm{peak}$ ($L_\sun$)\tablefootmark{b} \\ \cline{2-4} \cline{6-7}
 & Predict & No hist. & Hist. & & No hist. & Hist. & & &\\
\hline
s4n109   & 820 & 730 & 720 & 1.0 & 2660 & 5900 & 2.2 & \ldots\tablefootmark{c} & \ldots\tablefootmark{c}  \\
s75n89   &  90 & 170 & 290 & 1.7 &  330 &  740 & 2.3 & 143 &  5.8 \\
s156n98  & 110 & 220 & 360 & 1.6 &  240 &  480 & 2.0 &   8 & 13.6 \\
s257n132 & 220 & 240 & 280 & 1.2 &  540 &  750 & 1.4 &   4 &  8.8 \\
\hline
\hline                  
\end{tabular}
\tablefoot{
\tablefoottext{a}{The ratios are calculated between the measured extents with luminosity history, relative to no history.}
\tablefoottext{b}{Time since peak luminosity and the luminosity of the peak.}
\tablefoottext{c}{The luminosity of this object is highly variable and it is not possible to determine one single luminosity peak in the past that is responsible for the extended abundance profile. The luminosity of the object topped at 900\,$L_\sun$ \SI{650}{kyr} prior to the time of the snapshot.}
}
\end{table*}

Table~\ref{tbl:extents} records the extents of the four example objects, measured from both the fitted Gaussians and abundance profiles. The table also records the predicted extents calculated using the expression
\begin{equation}
  R_\mathrm{CO} = \SI{90}{AU} \left(\frac{L}{1\,L_\sun} \right)^{0.52}.
  \label{eq:extent}
\end{equation}
This expression is calculated by fitting Gaussians in the $(u,v)$-plane of a simple one-dimensional envelope model with a power law density profile, $\rho \propto r^{-p}$ with $p=1.5$, and a total mass of $1\,M_\sun$. The quantity $R_\mathrm{CO}$ is the Half Width Half Maximum of the fitted Gaussian (FWHM/2) and is reported in physical rather than angular units to make comparison with the abundance profiles easier. Table~\ref{tbl:extents} also gives the time interval since the peak luminosity as well as the peak luminosity itself.

For some objects it is not possible to measure the extent of \co accurately because most of the emission fall outside of the 15\,k$\uplambda$--50\,k$\uplambda$ baseline range used for fitting the Gaussians. An example of this is the object s4n109, whose luminosity is so high that most of the emission is at baselines <15\,k$\uplambda$. The high luminosity also means that even if such an object has undergone a luminosity burst in the past it will not be detected since the extended emission will be on even smaller baselines. The situation is similar for the objects s75n89 and s156n98, whose luminosities are so low that the half intensity point of the emission is at baselines significantly larger than 50\,k$\uplambda$. The difference relative to high-luminosity protostars is that, for low-luminosity protostars, bursts are easier to detect because the extended emission will be at shorter baselines covered by the interferometer.

For one of the objects, s75n89, the time interval since its peak luminosity is \SI{143}{kyr}, yet CO is still distributed over an extended area. This is a result of the density in the inner regions of the envelope being rather low (\SI{\sim e5}{cm^{-3}} corresponding to a freeze-out timescale of \SI{\sim 100}{kyr}). It is not clear that such a situation would be observed in nature since, at such long timescales, the dynamics of the system also becomes important. For the case at hand, the free-fall time for a particle sitting at a distance of \SI{700}{AU} from the protostar is \SI{\sim 10}{kyr} -- a factor of ten shorter than the time interval since the peak luminosity -- indicating that, after \SI{143}{kyr}, much of the gas-phase CO would have fallen towards the centre of the system.

Contamination from the ambient medium is an issue at the smallest baselines, as shown in Fig.~\ref{fig:uvamp} where the emission is seen to increase sharply at small baselines for three out of four systems (for the fourth system the luminosity is so high that most of the envelope emission is already at the small baselines). In nature, the situation is further aggravated by the presence of external heating through the ISRF and by low-density ``pre-depletion'' regions where CO has not yet frozen out, which may also increase the emission at the shortest baselines. This is an important consideration to take into account, when choosing how small baselines to include in the measurements.

\subsection{Comparison with observations}
\label{sec:comparison}

\begin{figure*}
\sidecaption
\includegraphics[width=12cm]{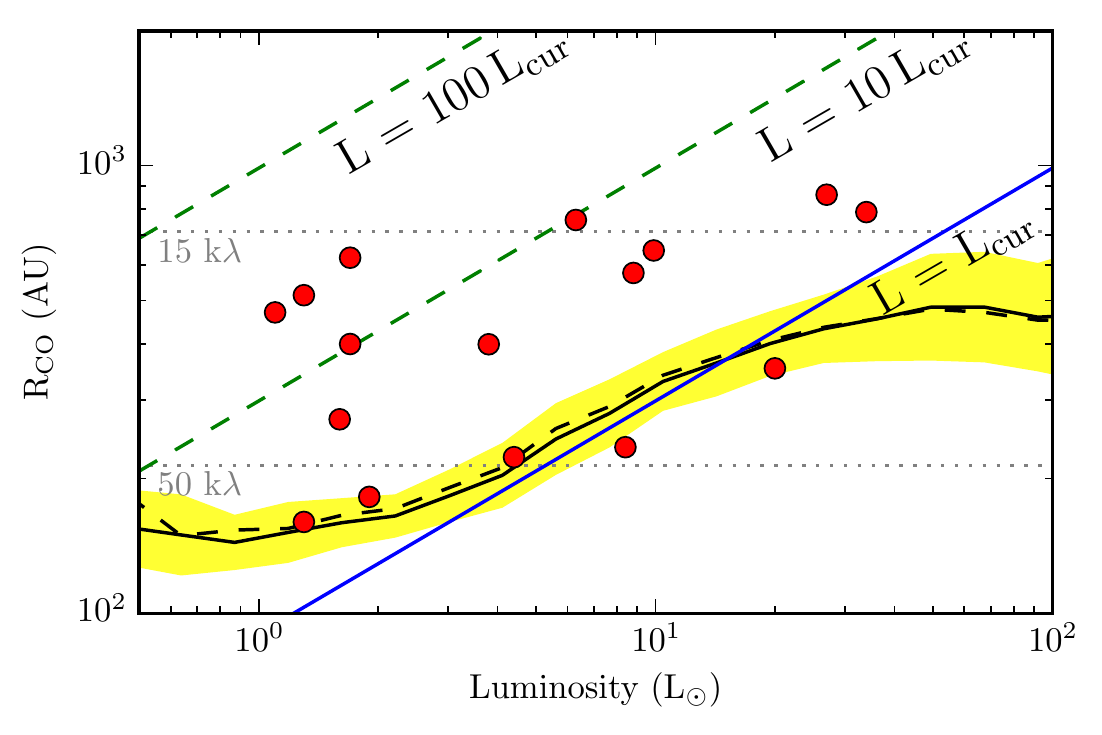}
   \caption{$R_\mathrm{CO}$ vs.\ luminosity. The black line is the measured extent from the simulation without any history, with the shaded area giving the one-sigma uncertainty. The black dashed line is the measured extent including luminosity history. The red points are the observations from \citet{Jorgensen:2015kz}. The blue line is the predicted extent given by Eq.~\eqref{eq:extent}. The dotted lines show the 15\,k$\uplambda$ and 50\,k$\uplambda$ baseline limits, assuming a source distance of \SI{235}{pc}.}
   \label{fig:rco}
\end{figure*}

Figure~\ref{fig:rco} shows $R_\mathrm{CO}$ as function of luminosity for the systems in the simulation. The black line in the figure is the binned median of $R_\mathrm{CO}$, measured for the protostars in the simulation, without including the luminosity history. The shaded area is the one-sigma standard deviation of $R_\mathrm{CO}$. At intermediate luminosities $R_\mathrm{CO}$ is seen to follow the blue line, which is the predicted extent given the current luminosity (see Eq.~\ref{eq:extent}). At both low and high luminosities the black curve flattens due to the 15 and~50\,$\mathrm{k}\uplambda$ baseline cut-offs. The measured extents do a little better than predicted for the low luminosities/small extents while for high luminosities/large extents they do a little worse. This is connected to the fact that the fitted Gaussians are very sensitive to the amplitude of the emission. For high luminosities, where most of the emission is at the short baselines, this often leads to an underestimation of the amplitude and consequently an underestimation of $R_\mathrm{CO}$. Conversely, for small extents, where it is easier to fit the amplitude, $R_\mathrm{CO}$ can be measured reliably even if it lies at baselines that are somewhat larger than those probed by the interferometer.

The red points in Fig.~\ref{fig:rco} are the protostars observed by \citet{Jorgensen:2015kz}, many of which have measured extents that are larger than can be explained by their current luminosity. Given that these are real observations there is an uncertainty associated with the measurements, originating from sources such as calibration, distance, and luminosity uncertainties. \citet{Jorgensen:2015kz} estimated that they could reliably detect extents corresponding to the luminosity being $\sim$5 times brighter than the current luminosity, which is the case for approximately half of the observed objects.

\begin{figure}
\includegraphics[width=\hsize]{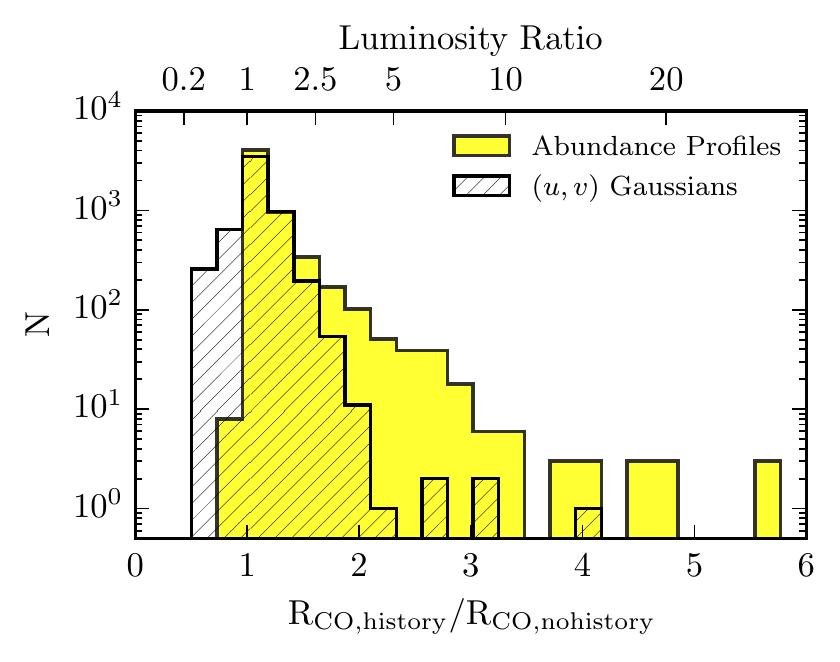}
\caption{Ratios of extents measured from the $(u,v)$-plane and from the abundance profiles. The lower axis show the measured ratios and the upper axis the corresponding luminosity ratio calculated using Eq.~\eqref{eq:extent}. $R_\mathrm{CO,no\,history}$ has been calculated using a wider $(u,v)$-coverage spanning from 5\,k$\uplambda$ to 200\,k$\uplambda$, to get a good reference value. $R_\mathrm{CO,history}$ has been calculated using the standard $(u,v)$-coverage.}
\label{fig:ratio}
\end{figure}

The black dashed line in Fig.~\ref{fig:rco} is the binned median of $R_\mathrm{CO}$ including luminosity history. The curve shows a marginally larger median value of $R_\mathrm{CO}$ relative to the situation with no history, however the increase is so slight that it is evident that only a small minority of systems are affected by the inclusion of luminosity history. Using Eq.~\eqref{eq:extent} as a reference \SI{\sim 1}{\percent} of the systems in the sample have a measured extent consistent with the protostar having been at least a factor of five brighter in the past.

Another way of illustrating the effects (or lack thereof) of including luminosity history is to calculate the ratio between the measured extents with and without luminosity history. A histogram of this is shown in Fig.~\ref{fig:ratio}, which also shows the same ratio calculated from the abundance profiles. Both histograms peak at a ratio of 1, with a tail towards higher ratios. The tail is significantly larger for the histogram of the abundance profiles, where \SI{2}{\percent} have a ratio consistent with the luminosity having been at least a factor of five brighter in the past. For the ratios measured by fitting Gaussians in the $(u,v)$-plane this is only true for six out of a total of more than \num{5800} cubes. There are several reasons why the ratios calculated from the abundance profiles have a larger tail towards high ratios -- the abundance profiles are, for example, not affected by baseline limits -- but the most important reason is that the radius measured for the abundance profile remain sensitive to an increased gas-phase abundance of CO for longer than the synthetic observables.

While differences in the physical structure of the envelopes are automatically accounted for in the measured extents potential variations of the sublimation temperature, $T_\mathrm{sub}$, are not automatically taken into account. Pure CO ice has a binding energy, $E_b/k_B$, in the range \SIrange[range-phrase = --]{855}{960}{K} \citep{Sandford:1993da,Bisschop:2006dc} corresponding to a sublimation temperature, $T_\mathrm{sub}$\,$\approx$\,\SI{20}{K}. For CO mixed with H$_2$O the binding energy is increased to $\approx$\SIrange[range-phrase = --]{1200}{1300}{K} \citep{Collings:2003kx,Noble:2012hq}, corresponding to $T_\mathrm{sub} \approx$\,\SI{28}{K}. A lower sublimation temperature would increase the predicted extent of CO at all luminosities, but would struggle to explain the compact emission measured towards some of the observed objects. A number of spatially unresolved observational studies have found CO to sublimate at \SIrange[range-phrase = --]{25}{40}{K} by coupling line observations of CO to simple parametric models of the envelopes with step functions describing the CO abundance \citep{Jorgensen:2004jh,Jorgensen:2005cs,Yldz:2013bt}. Despite the small samples and large uncertainties associated with these studies, they do suggest that the sublimation temperature of CO in cores is higher than that of pure CO ice, but also indicate that source-to-source variations may be important. The latter point especially may have important consequences for the interpretation of the observational results and underline the importance of both high quality laboratory measurements and large observational surveys dedicated to the study of the location of the CO ice lines.

\subsection{Baseline effects}
\label{sec:baseline}

The availability of baselines restrict the extents that can be measured with the interferometer. Baseline limits of~15\,k$\uplambda$ and~50\,k$\uplambda$ suggest that spatial extents between \ang{;;1.8} and~\ang{;;6.1} can be measured. A glance at the black line in Fig.~\ref{fig:rco} suggests that the actual limits are closer to \ang{;;1.3} and~\ang{;;4.0}. It is possible that the fully automatic fitting procedure applied here aggravate the situation somewhat and that better fits can be made for some of the sources by providing better starting guesses.

\begin{figure*}
\includegraphics[width=18cm]{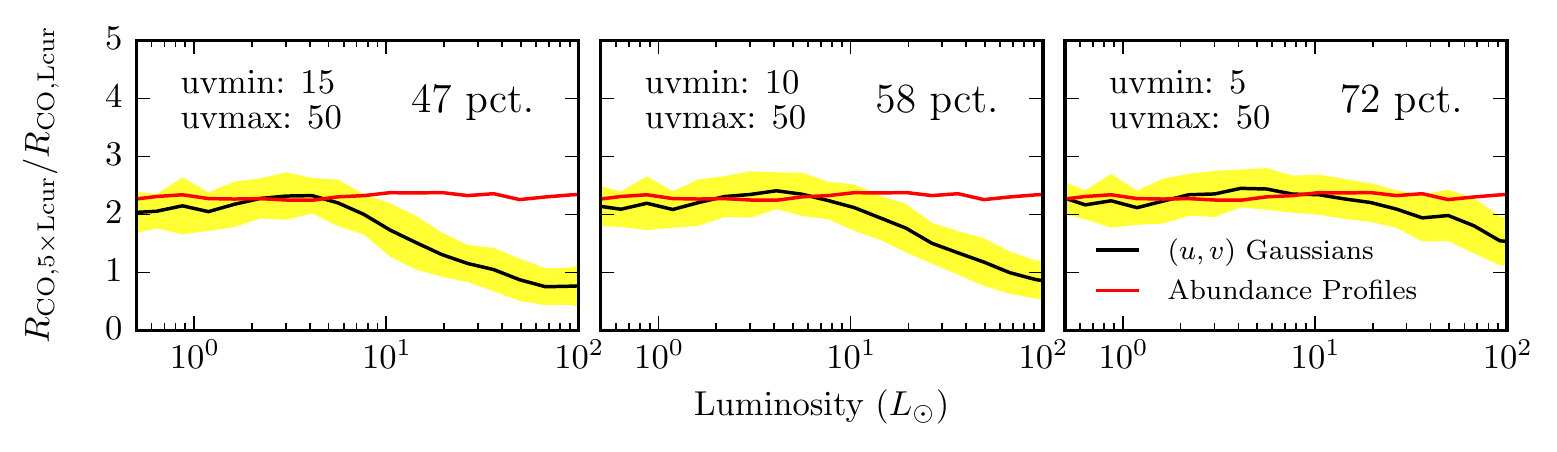}
   \caption{Ratios between measured extents as function of luminosity. The shaded area is the one-sigma uncertainty of the ratio measured from the $(u,v)$-plane. The $(u,v)$-range used for calculating the extent of $R_\mathrm{CO,5\times Lcur}$ is shown in the upper left of each panel. The number in the upper right corner of each panel indicate how large a fraction of the ratios measured from the $(u,v)$-plane are within \SI{20}{\percent} of the ratios measured from the abundance profiles.}
   \label{fig:ratio2}
\end{figure*}

To get a better grasp of the influence of the baseline limits on the results, we calculate a sample of synthetic line cubes where the luminosity of the central protostar has been enhanced by a factor of five, relative to the standard situation. The CO extents are measured in the cubes with enhanced luminosities for different baseline coverages and divided by the measured extents for the cubes with the non-enhanced luminosities to get the ratio. As in Fig.~\ref{fig:ratio} the extents of the reference cubes have been measured using a wide $(u,v)$-coverage spanning from 5\,k$\uplambda$ to 200\,k$\uplambda$, to provide a good reference value. A ratio is also calculated for the extents measured from the abundance profiles between the enhanced and non-enhanced luminosities.

Figure~\ref{fig:ratio2} shows the measured ratios as functions of luminosity. For a luminosity enhancement by a factor of five Eq.~\eqref{eq:extent} predicts a ratio of \num{2.3}, which is matched perfectly by the abundance profiles in the figure. The ratios calculated from the $(u,v)$-fits start declining at high luminosities because the emission is being spread out over larger spatial scales. Including smaller baselines in the fit is seen to delay the decline, making it possible to measure bursts at higher luminosities than normally possible. Access to longer baselines improve the measurements for small luminosities, but is not as crucial for the detection of bursts.

In general, one should be careful with including small baselines in the fits, however our results indicate that for luminosities $\gtrsim$\,5 it may be necessary in order to catch bursting sources. As seen in Fig.~\ref{eq:extent} the emission from the ambient medium typically only becomes important for baselines <10\,k$\uplambda$, as does the ISRF (\citealt{Jorgensen:2015kz}, Fig.~3). We therefore recommend 10\,k$\uplambda$ as a lower boundary when trying to measure large extents.

\section{Discussion}
\label{sec:discussion}

The spatial resolution of the simulation is \SI{50}{AU}, which is not enough to form circumstellar disks. The accretion rates onto the protostars in the simulation are therefore primarily modulated by variations in the infall of material from the large scales of the molecular cloud. Such variations, along with a gradual decrease in the mean accretion rate, are sufficient to reproduce the observed PLF \citep{Padoan:2014ho,Frimann:2015}, but, as the results of this paper demonstrate, they are not enough to reproduce the observed spread of \co extents seen in observations.

\begin{figure}
\includegraphics[width=\hsize]{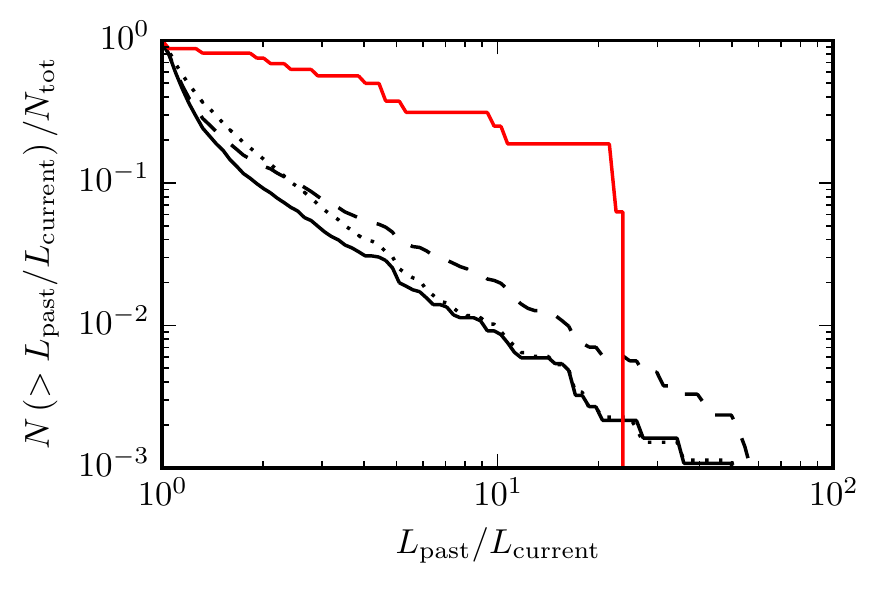}
\caption{Cumulative distribution of systems in the simulation that have had their luminosity increased by at least some factor, given by the horizontal axis, in the past. The black solid line is the distribution of the sample presented in Sect.~\ref{sec:targetselection}, while the red solid line is the distribution of the \citet{Jorgensen:2015kz} observations shown in Fig.~\ref{fig:rco}. The dotted line is for the same sample, but including binaries, while the dashed line include protostellar luminosities down to 0.05\,$L_\sun$. The lookback time is \SI{10}{kyr}.}
\label{fig:Lpast}
\end{figure}

To get a firm idea of how many systems with extended emission one can expect to find in the sample of protostars from the simulation, we analyse the luminosity time series of all the objects in the sample to see how large a fraction of the systems have had their luminosity enhanced by some factor in the past (black solid line in Fig.~\ref{fig:Lpast}; the red solid line shows the distribution of the real observations). We find that \SI{2}{\percent} of the systems have had their luminosity increased by at least a factor of five within the past \SI{10}{kyr} and \SI{1}{\percent} of the systems by at least a factor of 10. This is similar to the enhancements measured for the abundance profiles in Fig.~\ref{fig:ratio} and indicate that the discrepancy between the real and synthetic observations are due to the luminosity bursts in the simulation not being numerous or intensive enough.

It is possible that the selection criteria employed in Sect.~\ref{sec:targetselection} exclude important families of bursting protostars. Binary stars, for example, may be subject to increased variability since their accretion histories also depend on interactions with their companion. Redoing the analysis described above for the same sample of objects, but including binaries, reveal virtually no difference between the distributions (dotted line in Fig.~\ref{fig:Lpast}). We remark that, due to the relatively low spatial resolution in the simulation and the fact that protostars cannot form closer than \SI{400}{AU} to one another, the direct formation of close binaries is not modelled -- though close binaries resulting from dynamical interactions do occur -- and the analysis therefore does not rule out binarity as an important source of accretion variability.

Similarly, low-luminosity protostars may be subject to large fractional luminosity enhancements due to the low reference luminosity. Redoing the analysis yet again for the standard sample of protostars, but this time including objects with luminosities down to 0.05\,$L_\sun$, shows that the fraction of object at high luminosity ratios does increase significantly (dashed line in Fig.~\ref{fig:Lpast}), but still not enough to match the observed distribution.

The traditional view of FU Orionis type objects are that they are evolved objects, possibly with some remaining envelope material, with periodic luminosity outbursts driven by various types of disk instabilities \citep{Bell:1994cp,Armitage:2001jl,Vorobyov:2005kv,Zhu:2009fv,Martin:2011jz}. This view is supported by SED fitting of disk models to individual objects, such as FU Orionis itself \citep{Zhu:2007bi}. There is increasing evidence for the presence of disks, even around deeply embedded objects, both from an observational perspective \citep{Jorgensen:2009bx,Tobin:2012ee,Murillo:2013eq,Lindberg:2014bq,Tobin:2015fk}, as well as from numerical simulations \citep{Walch:2010ch,Walch:2012de,Seifried:2013hk,Nordlund+2014,Frimann:2015}, indicating that disk formation is ubiquitous at all stages of star formation. The lack of bursting sources in the simulation can be taken as indirect support both for the necessity of disks to mediate accretion, and of the existence of disks around the most deeply embedded protostars.

\section{Summary}
\label{sec:Summary}

This paper has presented an analysis of synthetic maps of the \co~$J$\,=\,2--1 line, targeting deeply embedded Class~0 and~I protostars, in a large MHD simulation of a molecular cloud. The primary goal of the paper has been to establish whether measurements of the spatial distribution of gas-phase \co around a sample of deeply embedded protostars, observed by \citet{Jorgensen:2015kz}, can be reproduced by the simulation. The main results of the paper are summarised as follows:
\begin{enumerate}
  \item One-dimensional Gaussian functions are fitted to the $(u,v)$-data to measure the spatial extent of \co. Fitting two-dimensional Gaussians reveals that, on average, the minor and major axes are within \SI{20}{\percent} of each other, indicating that the emission is mostly symmetric and in agreement with the real observations.
  \item CO abundance profiles are also analysed to measure the spatial extent of gas-phase CO independently of the fitted Gaussians. The two methods agree up to a roughly constant scale factor of \num[separate-uncertainty]{2.0 \pm 0.2}. This shows that the $(u,v)$-fitting method is capable of reliably measuring the CO extents.
  \item Given the adopted baseline limits of 15\,k$\uplambda$ and~50\,k$\uplambda$, we find that the spatial scales between which the extent of \co can be reliably measured are \ang{;;1.3} and~\ang{;;4.0}. This means that it is difficult to detect past luminosity bursts for high-luminosity sources because most emission will be at small baselines not included when measuring the extent. The adopted baseline limit of 15\,k$\uplambda$ was chosen to avoid issues with emission from the ambient medium; however, our analysis indicates that these issues only become important at baselines <10\,k$\uplambda$ and we therefore recommend using this as a limit when attempting to measure large extents.
  \item Approximately \SI{2}{\percent} of the systems in the simulation have \co distributed over an area that is extended enough to be consistent with the central protostar having been at least a factor of five brighter at some time in the past. For the systems observed by \citet{Jorgensen:2015kz} this fraction is approximately \SI{50}{\percent}. Because no disks are formed in the simulation, the protostellar luminosities wary owing to variations in the infall from large scales. Our results indicate that the presence of circumstellar disks, even in the embedded stages, is a necessary component to explain the wide distribution of extents seen in the observations.
  \item Even if the frequency and magnitude of episodic outbursts in the simulation are too small to explain the \co observations, it still reproduces the median and scatter of the PLF. This shows that the main physical attribute in setting up the PLF is not the presence of large and frequent accretion bursts, but rather the gradual decrease in accretion rate with time.
\end{enumerate}
The biggest uncertainty to the conclusions is whether the sublimation temperature of CO is subject to large source-to-source variations, which, if they are substantial, may explain most of the variations in the CO extents measured in the observations. This problem has to be investigated observationally, for example by observing the ice lines of other molecular species to see if they vary in phase with the CO ice lines. Even though the inclusion of disk physics is crucial for capturing the episodic accretion events, we note that it is equally important to ensure that the infall of material from large scales is realistic, otherwise the frequency and magnitude of the accretion bursts will not be correctly reproduced. Tentative results from high-resolution zoom-in simulations of individual protostars that include the larger scales indicate that the frequency of bursts increases once the circumstellar disk starts to become resolved \citep{Nordlund+2014}. Further high-resolution numerical follow-up studies, possibly including tracer particles to be able to follow the dynamical evolution of the systems, will help in deepening our understanding of the processes at play. This paper, which explores the influence of large-scale infall on the accretion histories of protostars, is an important step towards that goal.

\begin{acknowledgements}

We are grateful to the anonymous referee for the good comments, which improved the clarity of the paper. This research was supported by a Lundbeck Foundation Group Leader Fellowship to JKJ as well as the European Research Council (ERC) under the European Union's Horizon 2020 research and innovation programme (grant agreement No 646908) through ERC Consolidator Grant ``S4F''. PP acknowledges support by the Spanish MINECO under project AYA2014-57134-P. TH is supported by a Sapere Aude Starting Grant from the Danish Council for Independent Research. Research at Centre for Star and Planet Formation is funded by the Danish National Research Foundation and the University of Copenhagen's programme of excellence. We acknowledge PRACE for awarding us access to the computing resource CURIE based in France at CEA for carrying out part of the simulation. Archival storage and computing nodes at the University of Copenhagen HPC centre, funded with a research grant (VKR023406) from Villum Fonden, were used for carrying out part of the simulation and the post-processing.

\end{acknowledgements}

\bibliographystyle{aa} 
\bibliography{citations} 

\begin{thebibliography}{54}
\expandafter\ifx\csname natexlab\endcsname\relax\def\natexlab#1{#1}\fi

\bibitem[{Armitage {et~al.}(2001)Armitage, Livio, \& Pringle}]{Armitage:2001jl}
Armitage, P.~J., Livio, M., \& Pringle, J.~E. 2001, \mnras, 324, 705

\bibitem[{Audard {et~al.}(2014)Audard, {\'A}brah{\'a}m, Dunham, Green, Grosso,
  Hamaguchi, Kastner, K{\'o}sp{\'a}l, Lodato, Romanova, Skinner, Vorobyov, \&
  Zhu}]{Audard:2014gb}
Audard, M., {\'A}brah{\'a}m, P., Dunham, M.~M., {et~al.} 2014, in Protostars
  and Planets VI, ed. H.~{Beuther}, R.~S. {Klessen}, C.~P. {Dullemond}, \&
  T.~{Henning} (Tucson: University of Arizona Press), 387--410

\bibitem[{Bell \& Lin(1994)}]{Bell:1994cp}
Bell, K.~R. \& Lin, D. N.~C. 1994, \apj, 427, 987

\bibitem[{Bisschop {et~al.}(2006)Bisschop, Fraser, Oberg, van Dishoeck, \&
  Schlemmer}]{Bisschop:2006dc}
Bisschop, S.~E., Fraser, H.~J., Oberg, K.~I., van Dishoeck, E.~F., \&
  Schlemmer, S. 2006, Astronomy and Astrophysics, 449, 1297

\bibitem[{Bjorkman \& Wood(2001)}]{Bjorkman:2001du}
Bjorkman, J.~E. \& Wood, K. 2001, \apj, 554, 615

\bibitem[{Chen {et~al.}(1995)Chen, Myers, Ladd, \& Wood}]{Chen:1995eo}
Chen, H., Myers, P.~C., Ladd, E.~F., \& Wood, D. O.~S. 1995, \apj, 445, 377

\bibitem[{Collings {et~al.}(2003)Collings, Dever, Fraser, McCoustra, \&
  Williams}]{Collings:2003kx}
Collings, M.~P., Dever, J.~W., Fraser, H.~J., McCoustra, M. R.~S., \& Williams,
  D.~A. 2003, \apjl, 583, 1058

\bibitem[{D'Antona \& Mazzitelli(1997)}]{DAntona:1997vs}
D'Antona, F. \& Mazzitelli, I. 1997, \memsai, 68, 807

\bibitem[{Dullemond \& Dominik(2004)}]{Dullemond:2004iy}
Dullemond, C.~P. \& Dominik, C. 2004, \aap, 417, 159

\bibitem[{Dunham {et~al.}(2013)Dunham, Arce, Allen, Evans, Broekhoven-Fiene,
  Chapman, Cieza, Gutermuth, Harvey, Hatchell, Huard, Kirk, Matthews,
  Mer{\'\i}n, Miller, Peterson, \& Spezzi}]{Dunham:2013do}
Dunham, M.~M., Arce, H.~G., Allen, L.~E., {et~al.} 2013, \aj, 145, 94

\bibitem[{Dunham {et~al.}(2010)Dunham, Evans, Terebey, Dullemond, \&
  Young}]{Dunham:2010bx}
Dunham, M.~M., Evans, N.~J., Terebey, S., Dullemond, C.~P., \& Young, C.~H.
  2010, \apj, 710, 470

\bibitem[{Dunham \& Vorobyov(2012)}]{Dunham:2012ic}
Dunham, M.~M. \& Vorobyov, E.~I. 2012, \apj, 747, 52

\bibitem[{Evans {et~al.}(2009)Evans, Dunham, J{\o}rgensen, Enoch, Mer{\'\i}n,
  van Dishoeck, Alcal{\'a}, Myers, Stapelfeldt, Huard, Allen, Harvey, van
  Kempen, Blake, Koerner, Mundy, Padgett, \& Sargent}]{Evans:2009bk}
Evans, N.~J., Dunham, M.~M., J{\o}rgensen, J.~K., {et~al.} 2009, \apjs, 181,
  321

\bibitem[{{Frimann} {et~al.}(2015){Frimann}, {J{\o}rgensen}, \&
  {Haugb{\o}lle}}]{Frimann:2015}
{Frimann}, S., {J{\o}rgensen}, J.~K., \& {Haugb{\o}lle}, T. 2015, ArXiv
  e-prints [\eprint[arXiv]{1510.07827}]

\bibitem[{Herbig(1977)}]{Herbig:1977gf}
Herbig, G.~H. 1977, \apj, 217, 693

\bibitem[{Johnstone {et~al.}(2013)Johnstone, Hendricks, Herczeg, \&
  Bruderer}]{Johnstone:2013je}
Johnstone, D., Hendricks, B., Herczeg, G.~J., \& Bruderer, S. 2013, \apj, 765,
  133

\bibitem[{J{\o}rgensen(2004)}]{Jorgensen:2004jh}
J{\o}rgensen, J.~K. 2004, \aap, 424, 589

\bibitem[{J{\o}rgensen {et~al.}(2002)J{\o}rgensen, Sch{\"o}ier, \& van
  Dishoeck}]{Jorgensen:2002ci}
J{\o}rgensen, J.~K., Sch{\"o}ier, F.~L., \& van Dishoeck, E.~F. 2002, \aap,
  389, 908

\bibitem[{J{\o}rgensen {et~al.}(2005)J{\o}rgensen, Sch{\"o}ier, \& van
  Dishoeck}]{Jorgensen:2005cs}
J{\o}rgensen, J.~K., Sch{\"o}ier, F.~L., \& van Dishoeck, E.~F. 2005, \aap,
  435, 177

\bibitem[{J{\o}rgensen {et~al.}(2009)J{\o}rgensen, van Dishoeck, Visser,
  Bourke, Wilner, Lommen, Hogerheijde, \& Myers}]{Jorgensen:2009bx}
J{\o}rgensen, J.~K., van Dishoeck, E.~F., Visser, R., {et~al.} 2009, \aap, 507,
  861

\bibitem[{J{\o}rgensen {et~al.}(2013)J{\o}rgensen, Visser, Sakai, Bergin,
  Brinch, Harsono, Lindberg, van Dishoeck, Yamamoto, Bisschop, \&
  Persson}]{Jorgensen:2013ej}
J{\o}rgensen, J.~K., Visser, R., Sakai, N., {et~al.} 2013, \apjl, 779, L22

\bibitem[{J{\o}rgensen {et~al.}(2015)J{\o}rgensen, Visser, Williams, \&
  Bergin}]{Jorgensen:2015kz}
J{\o}rgensen, J.~K., Visser, R., Williams, J.~P., \& Bergin, E.~A. 2015, \aap,
  579, A23

\bibitem[{Kenyon {et~al.}(1990)Kenyon, Hartmann, Strom, \&
  Strom}]{Kenyon:1990ki}
Kenyon, S.~J., Hartmann, L.~W., Strom, K.~M., \& Strom, S.~E. 1990, \aj, 99,
  869

\bibitem[{Kryukova {et~al.}(2012)Kryukova, Megeath, Gutermuth, Pipher, Allen,
  Allen, Myers, \& Muzerolle}]{Kryukova:2012ea}
Kryukova, E., Megeath, S.~T., Gutermuth, R.~A., {et~al.} 2012, \aj, 144, 31

\bibitem[{Lindberg {et~al.}(2014)Lindberg, J{\o}rgensen, Brinch, Haugb{\o}lle,
  Bergin, Harsono, Persson, Visser, \& Yamamoto}]{Lindberg:2014bq}
Lindberg, J.~E., J{\o}rgensen, J.~K., Brinch, C., {et~al.} 2014, \aap, 566, 74

\bibitem[{Lomax {et~al.}(2014)Lomax, Whitworth, Hubber, Stamatellos, \&
  Walch}]{Lomax:2014bw}
Lomax, O., Whitworth, A.~P., Hubber, D.~A., Stamatellos, D., \& Walch, S. 2014,
  \mnras, 439, 3039

\bibitem[{Martin \& Lubow(2011)}]{Martin:2011jz}
Martin, R.~G. \& Lubow, S.~H. 2011, \apj, 740, L6

\bibitem[{Murillo {et~al.}(2013)Murillo, Lai, Bruderer, Harsono, \& van
  Dishoeck}]{Murillo:2013eq}
Murillo, N.~M., Lai, S.-P., Bruderer, S., Harsono, D., \& van Dishoeck, E.~F.
  2013, \aap, 560, 103

\bibitem[{Myers(2012)}]{Myers:2012ea}
Myers, P.~C. 2012, \apj, 752, 9

\bibitem[{Myers \& Ladd(1993)}]{Myers:1993en}
Myers, P.~C. \& Ladd, E.~F. 1993, \apj, 413, L47

\bibitem[{Noble {et~al.}(2012)Noble, Congiu, Dulieu, \& Fraser}]{Noble:2012hq}
Noble, J.~A., Congiu, E., Dulieu, F., \& Fraser, H.~J. 2012, \mnras, 421, 768

\bibitem[{{Nordlund} {et~al.}(2014){Nordlund}, {Haugb{\o}lle}, {K{\"u}ffmeier},
  {Padoan}, \& {Vasileiades}}]{Nordlund+2014}
{Nordlund}, {\AA}., {Haugb{\o}lle}, T., {K{\"u}ffmeier}, M., {Padoan}, P., \&
  {Vasileiades}, A. 2014, in IAU Symposium, Vol. 299, IAU Symposium, ed.
  M.~{Booth}, B.~C. {Matthews}, \& J.~R. {Graham}, 131--135

\bibitem[{Offner \& McKee(2011)}]{Offner:2011ex}
Offner, S. S.~R. \& McKee, C.~F. 2011, \apj, 736, 53

\bibitem[{Ossenkopf \& Henning(1994)}]{Ossenkopf:1994tq}
Ossenkopf, V. \& Henning, T. 1994, \aap, 291, 943

\bibitem[{Padoan {et~al.}(2014)Padoan, Haugb{\o}lle, \&
  Nordlund}]{Padoan:2014ho}
Padoan, P., Haugb{\o}lle, T., \& Nordlund, {\AA}. 2014, \apj, 797, 32

\bibitem[{Rodgers \& Charnley(2003)}]{Rodgers:2003dg}
Rodgers, S.~D. \& Charnley, S.~B. 2003, \apj, 585, 355

\bibitem[{Safron {et~al.}(2015)Safron, Fischer, Megeath, Furlan, Stutz, Stanke,
  Billot, Rebull, Tobin, Ali, Allen, Booker, Watson, \& Wilson}]{Safron:2015ew}
Safron, E.~J., Fischer, W.~J., Megeath, S.~T., {et~al.} 2015, \apj, 800, L5

\bibitem[{Sandford \& Allamandola(1993)}]{Sandford:1993da}
Sandford, S.~A. \& Allamandola, L.~J. 1993, \apj, 417, 815

\bibitem[{Seifried {et~al.}(2013)Seifried, Banerjee, Pudritz, \&
  Klessen}]{Seifried:2013hk}
Seifried, D., Banerjee, R., Pudritz, R.~E., \& Klessen, R.~S. 2013, \mnras,
  432, 3320

\bibitem[{Shirley {et~al.}(2002)Shirley, Evans, \& Rawlings}]{Shirley:2002co}
Shirley, Y.~L., Evans, N.~J., \& Rawlings, J. M.~C. 2002, \apj, 575, 337

\bibitem[{Shirley {et~al.}(2011)Shirley, Huard, Pontoppidan, Wilner, Stutz,
  Bieging, \& Evans}]{Shirley:2011jz}
Shirley, Y.~L., Huard, T.~L., Pontoppidan, K.~M., {et~al.} 2011, \apj, 728, 143

\bibitem[{Teyssier(2002)}]{Teyssier:2002fj}
Teyssier, R. 2002, \aap, 385, 337

\bibitem[{Tobin {et~al.}(2012)Tobin, Hartmann, Chiang, Wilner, Looney, Loinard,
  Calvet, \& D'Alessio}]{Tobin:2012ee}
Tobin, J.~J., Hartmann, L., Chiang, H.-F., {et~al.} 2012, \nat, 492, 83

\bibitem[{Tobin {et~al.}(2015)Tobin, Looney, Wilner, Kwon, Chandler, Bourke,
  Loinard, Chiang, Schnee, \& Chen}]{Tobin:2015fk}
Tobin, J.~J., Looney, L.~W., Wilner, D.~J., {et~al.} 2015, \apj, 805, 125

\bibitem[{Visser \& Bergin(2012)}]{Visser:2012dp}
Visser, R. \& Bergin, E.~A. 2012, \apjl, 754, L18

\bibitem[{Visser {et~al.}(2015)Visser, Bergin, \& J{\o}rgensen}]{Visser:2015ew}
Visser, R., Bergin, E.~A., \& J{\o}rgensen, J.~K. 2015, \aap, 577, A102

\bibitem[{Vorobyov {et~al.}(2013)Vorobyov, Baraffe, Harries, \&
  Chabrier}]{Vorobyov:2013kq}
Vorobyov, E.~I., Baraffe, I., Harries, T., \& Chabrier, G. 2013, \aap, 557, A35

\bibitem[{Vorobyov \& Basu(2005)}]{Vorobyov:2005kv}
Vorobyov, E.~I. \& Basu, S. 2005, \apj, 633, L137

\bibitem[{Walch {et~al.}(2010)Walch, Naab, Whitworth, Burkert, \&
  Gritschneder}]{Walch:2010ch}
Walch, S., Naab, T., Whitworth, A., Burkert, A., \& Gritschneder, M. 2010,
  \mnras, 402, 2253

\bibitem[{Walch {et~al.}(2012)Walch, Whitworth, \& Girichidis}]{Walch:2012de}
Walch, S., Whitworth, A.~P., \& Girichidis, P. 2012, \mnras, 419, 760

\bibitem[{Young \& Evans(2005)}]{Young:2005ic}
Young, C.~H. \& Evans, N.~J. 2005, \apj, 627, 293

\bibitem[{Yıldız {et~al.}(2013)Yıldız, Kristensen, van Dishoeck, San
  Jose-Garcia, Karska, Harsono, Tafalla, Fuente, Visser, J{\o}rgensen, \&
  Hogerheijde}]{Yldz:2013bt}
Yıldız, U.~A., Kristensen, L.~E., van Dishoeck, E.~F., {et~al.} 2013, \aap,
  556, A89

\bibitem[{Zhu {et~al.}(2007)Zhu, Hartmann, Calvet, Hernandez, Muzerolle, \&
  Tannirkulam}]{Zhu:2007bi}
Zhu, Z., Hartmann, L., Calvet, N., {et~al.} 2007, \apj, 669, 483

\bibitem[{Zhu {et~al.}(2009)Zhu, Hartmann, \& Gammie}]{Zhu:2009fv}
Zhu, Z., Hartmann, L., \& Gammie, C. 2009, \apj, 694, 1045

\end{thebibliography}

\end{document}